\documentclass{article}
\usepackage{spconf,amsmath,graphicx}
\usepackage{booktabs,url}

\newcommand{\Fig}[1]{\textbf{Fig. \ref{fig:#1}}} 
\newcommand{\Eq}[1]{Eq. (\ref{eq:#1})} 
\newcommand{\Sec}[1]{\textbf{Section \ref{sec:#1}}} 

\newcommand{\Norm}[1]{ {\mathcal N}\left( #1 \right) } 
 
\renewcommand{\Vec}[1]{\textrm{\boldmath $#1$}} 
\newcommand{\Argmax}{\mathop {{\rm argmax}}}
\newcommand{\Argmin}{\mathop {{\rm argmin}}}
\newcommand{\pt}[1]{\left(#1\right)} 
\newcommand{\br}[1]{\left[#1\right]} 
\newcommand{\x}{ \Vec{x} } 
\newcommand{\z}{ \Vec{z} } 
\newcommand{\hatx}{ \Vec{\hat x} } 

\newcommand{\drawfig}[4]{ 
  \begin{figure}[#1]
  \centering \vspace{-0mm}
  \includegraphics[width=#2,clip]{#3.eps} \vspace{-3mm} 
  \caption{#4} \vspace{-4mm}
  \label{fig:#3}
  \end{figure}
}
\newcommand{\drawfigwide}[4]{ 
  \begin{figure*}[#1]
  \centering \vspace{-0mm}
  \includegraphics[width=#2,clip]{#3.eps} \vspace{-3mm} 
  \caption{#4} \vspace{-4mm}
  \label{fig:#3}
  \end{figure*}
}

\title{HumanGAN: generative adversarial network with human-based discriminator and its evaluation in speech perception modeling}

\makeatletter
\def\name#1{\gdef\@name{#1\\}}
\makeatother
\name{{\em Kazuki Fujii$^{1,2}$, Yuki Saito$^{1}$, Shinnosuke Takamichi$^{1}$, Yukino Baba$^{3}$, and Hiroshi Saruwatari$^{1}$}}
\address{
    $^1$ National Institute of Technology, Tokuyama College, Japan. \\    
    $^2$ Graduate School of Information Science and Technology, The University of Tokyo, Japan. \\
    $^3$ Faculty of Engineering, Information and Systems, University of Tsukuba, Japan.
}

\begin{document}
\ninept
\maketitle

\setlength{\abovedisplayskip}{3pt} 
\setlength{\belowdisplayskip}{3pt} 
\allowdisplaybreaks

\begin{abstract}
    We propose the HumanGAN, a generative adversarial network (GAN) incorporating human perception as a discriminator. A basic GAN trains a generator to represent a real-data distribution by fooling the discriminator that distinguishes real and generated data. Therefore, the basic GAN cannot represent the outside of a real-data distribution. In the case of speech perception, humans can recognize not only human voices but also processed (i.e., a non-existent human) voices as human voice. Such a human-acceptable distribution is typically wider than a real-data one and cannot be modeled by the basic GAN. To model the human-acceptable distribution, we formulate a backpropagation-based generator training algorithm by regarding human perception as a black-boxed discriminator. The training efficiently iterates generator training by using a computer and discrimination by crowdsourcing. We evaluate our HumanGAN in speech naturalness modeling and demonstrate that it can represent a human-acceptable distribution that is wider than a real-data distribution. 
\end{abstract}


\vspace{-1mm}
\section{Introduction} \vspace{-1mm}
    Generative models in machine learning have contributed strongly to media (e.g., speech)-related research. In particular, deep neural network (DNN)-based generative models, which are called ``deep generative models,'' can model a complicated data distribution thanks to the nonlinear transformation of DNNs. A generative adversarial network (GAN)~\cite{goodfellow14gan} is one of the most promising approaches in learning deep generative models. The basic GAN framework includes not only a generator but also a discriminator. The basic GAN trains the discriminator by distinguishing real and generated data and separately trains the generator by fooling the discriminator. By iterating the training steps, the generator can randomly generate samples that follow a real-data distribution, i.e., training-data distribution. Many studies on speech~\cite{saito18advss}, image~\cite{mirza15conditionalgan}, and language~\cite{yu19seqgan} have successfully applied the basic GAN framework, and more studies are expected to appear in speech-related research.
    
    However, the basic GAN represents only a real-data distribution\footnote{This problem is common among generative models, but this paper deals only with GAN.}. Human speech perception can accept a deviation from real speech. For instance, humans can recognize both human speech and processed (i.e., non-existent human) speech as human voice. In this paper, we call such a human-acceptable distribution the \textit{perception distribution} and establish a GAN framework that can represent it. If the perception distribution covers wider ranges than a real-data distribution, a basic GAN trained by fooling the discriminator could never represent the perception distribution. Also, collecting a large amount of real data for training the basic GAN would never solve this problem.
    
    In this paper, we propose the \textit{HumanGAN} trained by fooling human perception to model a human perception distribution. The HumanGAN regards a crowd as a black-boxed system that outputs a difference of posterior probabilities (i.e., to what degree do humans accept the data) given generated data. We formulate a backpropagation-based generator training algorithm utilizing an optimization algorithm for the black-boxed system. \Fig{overview} shows a comparison of the basic GAN and proposed HumanGAN. Whereas a generator of the basic GAN fools a DNN-based discriminator, that of the proposed HumanGAN fools perceptual evaluation performed by the crowdworkers of a crowdsourcing service. In this paper, we evaluate the HumanGAN in the modeling of speech naturalness, i.e., to what degree can humans accept synthesized speech as human voice. The experimental results demonstrate that 1) the perception distribution covers a wider range than the real-data distribution, and 2) the proposed HumanGAN accurately models the perception distribution, which the basic GAN cannot.

    \drawfig{t}{0.98\linewidth}{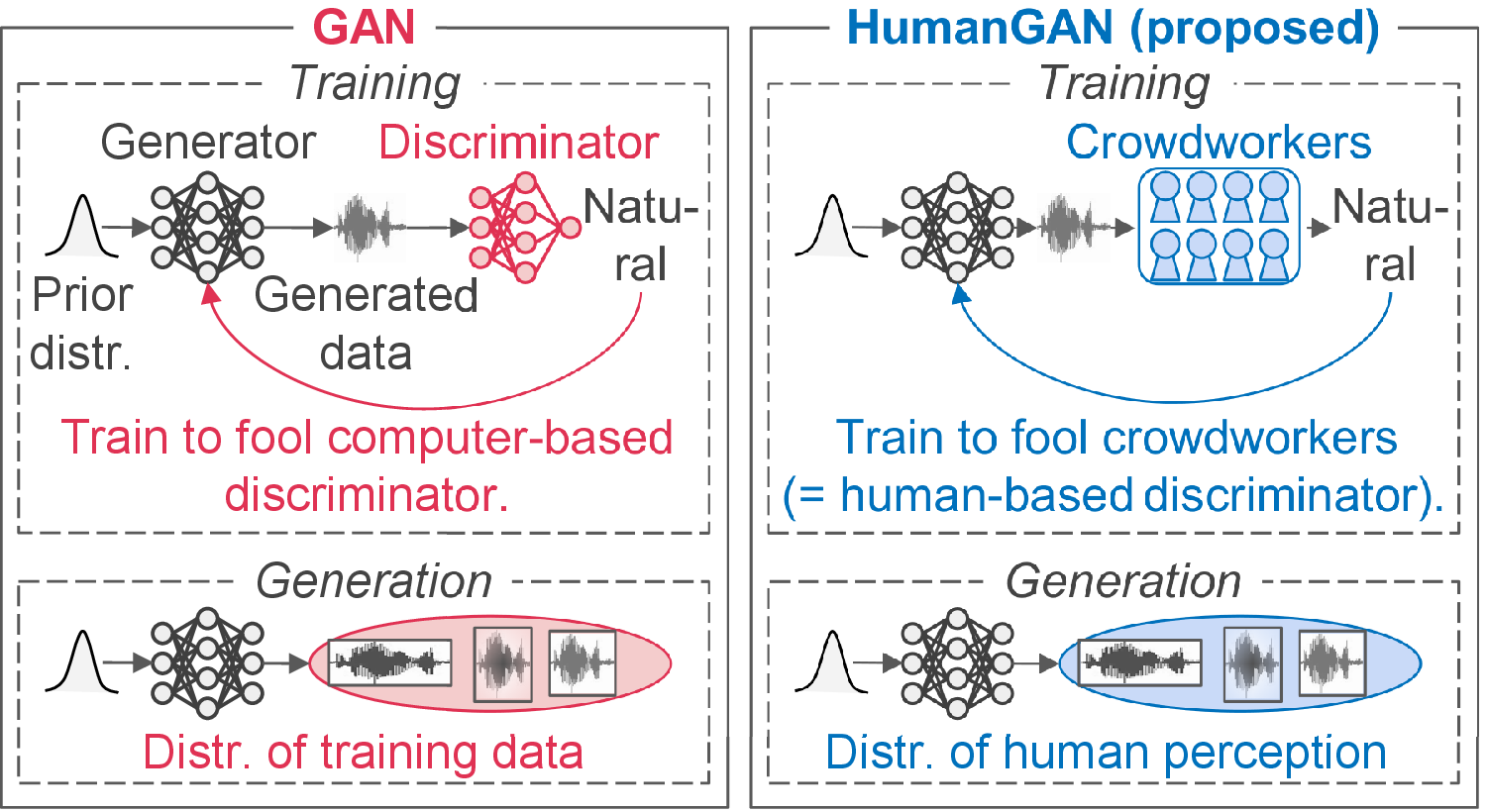}
    {Comparison of basic GAN and proposed HumanGAN. Basic GAN trains generator by fooling DNN-based discriminator (i.e., computer-based discriminator), and generator finally represents training-data distribution. In comparison, HumanGAN trains generator by fooling crowdworkers' perceptual evaluation (i.e., human-based discriminator), and generator finally represents humans' perception distribution.}

\vspace{-1mm}
\section{Basic GAN using DNN-based discriminator} \vspace{-1mm}
    The aim of the basic GAN is to match a generated-data distribution and real-data distribution. To achieve this, the basic GAN uses not only a generator $G\pt{\cdot}$ that randomly generates data but also a discriminator $D\pt{\cdot}$ that distinguishes real and generated data. Here, let real data be $\x = \br{\x_1, \cdots, \x_n, \cdots, \x_N}$, and $N$ be the number of data. The generator $G\pt{\cdot}$ transforms prior noise $\z = \br{\z_1, \cdots, \z_n, \cdots, \z_N}$ into generated data $\hatx = \br{\hatx_1, \cdots, \hatx_n, \cdots, \hatx_N}$. The prior noise follows a known probability distribution, e.g., a uniform distribution. The input of the discriminator $D\pt{\cdot}$ is real data $\x_n$ or generated data $\hatx_n$, and the $D\pt{\cdot}$ outputs a posterior probability that the input is real data. An objective function in training is formulated as
        \begin{align}
            V\pt{G, D} &= \sum\limits_{n=1}^N \log D\pt{\x_n} + \sum\limits_{n=1}^N \log \pt{1 - D\pt{G\pt{\z_n}}}.
            \label{eq:gan_loss}
        \end{align}
    The generator $G\pt{\cdot}$ and discriminator $D\pt{\cdot}$ are trained one by one. The following sections describe these training steps.

    \subsection{Generator training}
        The generator training step minimizes \Eq{gan_loss} to estimate the model parameters of $G\pt{\cdot}$. The model parameter $\theta_{\rm G}$ is estimated:
            \begin{align}
                \theta_{\rm G} = \Argmin_{\theta_{\rm G}} \sum\limits_{n=1}^N \log \pt{1 - D\pt{G\pt{\z_n}}}.
                \label{eq:gan_loss_g}
            \end{align}
        Namely, $G\pt{\cdot}$ is trained to let $D\pt{\cdot}$ recognize the generated data as ``real.'' In general, all computation processes are differentiable, and $\theta_{\rm G}$ is iteratively estimated by using the backpropagation algorithm.

    \subsection{Discriminator training}
        The discriminator training step maximizes \Eq{gan_loss} to estimate the model parameters of $D\pt{\cdot}$. The model parameter $\theta_{\rm D}$ is estimated:
            \begin{align}
                \theta_{\rm D} = \Argmax_{\theta_{\rm D}} V\pt{G, D}.
                \label{eq:gan_loss_d}
            \end{align}
    
    \subsection{Problem}
        When the generator is accurately trained, it can randomly sample data that follows a real-data distribution. However, when the perception distribution covers ranges wider than the real-data distribution, the basic GAN cannot represent the ranges. We show such an actual example in speech perception in \Sec{diff}.

\vspace{-1mm}
\section{HumanGAN using human-based discriminator} \vspace{-1mm}
        
        \drawfig{t}{0.85\linewidth}{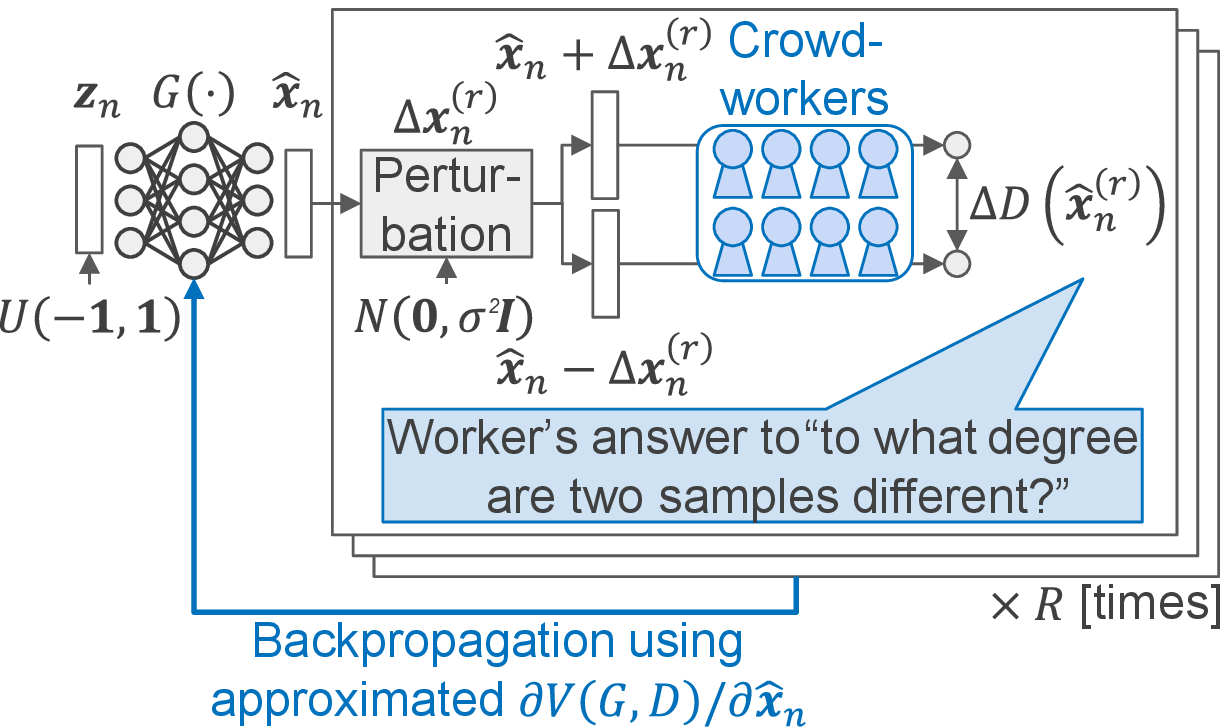}
        {Generator training procedure of proposed HumanGAN. Crowdworkers state a perceptual difference (i.e., difference of posterior probabilities) of two perturbed samples. Answer and perturbation are used for backpropagation to train generator.}
        
    We propose the HumanGAN to represent humans' perception distribution. As shown in \Fig{overview}, the HumanGAN replaces the DNN-based discriminator of the basic GAN with humans' perceptual evaluation. The use of a DNN-based generator and prior noise are common between the basic GAN and HumanGAN. However, whereas the basic GAN trains the generator by fooling the DNN-based discriminator, the HumanGAN trains one by fooling humans' perceptual evaluation. Here, we re-define $D\pt{\cdot}$ as a perception model that represents humans' perceptual evaluation. An input of the $D\pt{\cdot}$ is $\hatx_n$ generated from $G\pt{\cdot}$, and the $D\pt{\cdot}$ outputs a posterior probability about ``to what degree is the input perceptually acceptable'' from 0 to 1. The objective function during training is as follows.

        \begin{align}
            V\pt{G, D} = \sum\limits_{n=1}^N D\pt{G\pt{\z_n}}.
            \label{eq:humangan_loss}
        \end{align}
    As described in this equation, the HumanGAN never uses real-data in training.

    \subsection{Generator training}\label{sec:humangan_train}
        A model parameter $\theta_{\rm G}$ of $G\pt{\cdot}$ is estimated by maximizing \Eq{humangan_loss}. In this paper, we formulate a gradient-based iterative method. Namely, $\theta_{\rm G}$ is iteratively updated as follows.
            \begin{align}
                \theta_{\rm G} \leftarrow \theta_{\rm G} + \alpha \frac{\partial V\pt{G, D}}{\partial \theta_{\rm G}},
            \end{align}
        where $\alpha$ is the learning coefficient. The second term $\partial V\pt{G, D} / \partial \theta_{\rm G}$ is decomposed into a product of $\partial V\pt{G, D} / \partial \hatx$ and $\partial \hatx / \partial \theta_{\rm G}$. In the basic GAN, $\partial V\pt{G, D} / \partial \theta_{\rm G}$ is estimated by using the standard backpropagation algorithm since the computation processes of both $G\pt{\cdot}$ and $D\pt{\cdot}$ are differentiable. However, $D\pt{\cdot}$ is not differentiable in the HumanGAN. Therefore, we cannot estimate $\partial V\pt{G, D} / \partial \hatx$ and cannot use the standard backpropagation algorithm. In this paper, we regard humans as a black-boxed system that outputs a difference of the posterior probabilities of generated data and estimate $\partial V\pt{G, D} / \partial \hatx$ on the basis of an optimization algorithm for black-boxed systems. We propose a training algorithm that uses natural evolution strategies (NES)~\cite{ilyas18blackboxlimitedquery} that approximates gradients by using data perturbations. \Fig{training} shows the training procedure.
        
        First, a small perturbation $\Delta \x_{n}^{(r)}$ is randomly generated from a multivariate Gaussian distribution $\Norm{\Vec{0}, \sigma^2\Vec{I}}$. $r$ is the perturbation index ($1 \leq r \leq R$). $\sigma$ is a constant value of standard deviation. $\Vec{I}$ is the identity matrix. Then, a human is presented with two perturbed data $\left\{ \hatx_n + \Delta \x_{n}^{(r)}, \hatx_n - \Delta \x_{n}^{(r)} \right\}$ and evaluates the difference of their posterior probabilities: 
            \begin{align}
                \Delta D(\hatx_n^{(r)}) \equiv D\pt{\hatx_n + \Delta \x_{n}^{(r)}} - D\pt{\hatx_n - \Delta \x_{n}^{(r)}}.
            \end{align}
        $\Delta D(\hatx_n^{(r)})$ ranges from $-1$ to $1$. For example, we expect that a human will return $\Delta D(\hatx_n^{(r)})=1$ when he/she perceives that $D(\hatx_n + \Delta \x_{n}^{(r)})$ is greatly higher than $D(\hatx_n - \Delta \x_{n}^{(r)})$. This perturbation and evaluation are iterated $R$ times for one $\hatx_n$. Finally, these processes are iterated for $N$ generated data.

        $\partial V\pt{G, D} / \partial \hatx$ for the backpropagation algorithm is approximated as~\cite{ilyas18blackboxlimitedquery}
            \begin{align}
                \frac{\partial V\pt{G, D}}{\partial \hatx}  &= \br{\frac{\partial V\pt{G, D}}{\partial \hatx_1}, \cdots, \frac{\partial V\pt{G, D}}{\partial \hatx_n}, \cdots, \frac{\partial V\pt{G, D}}{\partial \hatx_N}}, \\
                \frac{\partial V\pt{G, D}}{\partial \hatx_n}&= \frac{1}{2\sigma R} \sum\limits_{r=1}^{R} 
                \Delta D(\hatx_n^{(r)}) \cdot \Delta \x_{n}^{(r)}. 
            \end{align}
        Here, we summarize the number of queries to humans for training the generator. Since the perturbation is performed $R$ times for each generated data, the total number of queries is a product of the number of training iterations, the number of generated data $N$, and the number of perturbations $R$.  

    \subsection{Discussion}
        Our work utilizes the idea of \textit{human computation}~\cite{humancomputationsurvey} in which a machine performs its function by outsourcing inner steps to humans, and the HumanGAN is regarded as a human-in-the-loop machine learning technique. Some existing studies made humans have a function and integrated it into DNN training. Sakata et al.~\cite{sakata19crownn} regarded humans as a feature extractor and improved the performance of a multi-class classifier. Also, Jaques et al.~\cite{jaques18learningsocialaware} regarded humans as an auxiliary classifier and improved performances of a generator. From these viewpoints, our work regards humans as a system that outputs a difference of posterior probabilities and represents humans' perceptual distribution.
        
        Koyama et al.'s work~\cite{koyama14crowdanalysis} is related work from the viewpoint of exploring humans' perception. However, their aim is completely different from ours. Koyama et al. made humans score a perceptual difference of two data and aimed at visualizing the perception distribution by interpolating the scores. In comparison, our work makes humans score the perceptual difference of a perturbation and aims at training a generator that represents the perception distribution.  
            
        Natural language-based image generation and editing~\cite{mansimov2016genimgfrmcaption,shinagawa19imagemanipulation} are related work from the viewpoint of modeling a data distribution that corresponds to human intention. However, these methods only estimate a conditional distribution within a real-data distribution. In comparison, our work estimates a perception distribution even if it is wider than the real-data distribution.

        The basic GAN utilizes only a computer-based discriminator, and the HumanGAN utilizes only a human-based discriminator. One part of our future work involves the use of a human-computer discriminator, i.e., the generator is trained by using both computer-based and human-based discriminators. Related studies have proposed active learning algorithms that use the basic GAN~\cite{zhu17adversarialactivelearning,deng18adversarialactivelabeling}. These studies are currently limited to real-data distribution modeling but have the potential to be applied to perception distribution modeling.
            
        Data imbalance in training data is a general problem in machine learning. Since the HumanGAN does not require real data for the training, it has the possibility of relaxing the problem.

    \subsection{Practical limitation}
        Here, we list empirically found limitations of the HumanGAN. 
    
        As described in \Eq{humangan_loss}, the generator of the HumanGAN is trained by maximizing the posterior probability. Therefore, a mode collapse problem~\cite{goodfellow16gantutorial} is caused when data generated from the initial generator is distributed near the mode of the perception distribution. Also, a gradient vanishing problem is caused when generated data is distributed far from the perception distribution. In the evaluation in \Sec{learnmodel}, we used real data for initializing the generator.
        
        The setting of $\sigma$ in the NES algorithm changes the degree of perturbation. When $\sigma$ is very small, the discriminator of the HumanGAN, i.e., crowdworkers, cannot find a perceptual difference, and it makes gradients vanish. When $\sigma$ is very large, the gradient becomes inaccurate. In the preliminary experiments of the following evaluation, we used several settings for $\sigma$ and finally determined one value.
        
        As described in \Sec{humangan_train}, the number of queries to crowdworkers is a product of the number of training iterations, the number of data $N$, and the number of perturbations $R$. The number of queries increases the monetary and time costs for training. In the following evaluation, we used a small generator, a small number of generated data, and low-dimensional data to limit the costs.

\vspace{-1mm}
\section{Experimental evaluation} \vspace{-1mm}
    \subsection{Experimental setup} \label{sec:expcond}
        \drawfig{t}{0.60\linewidth}{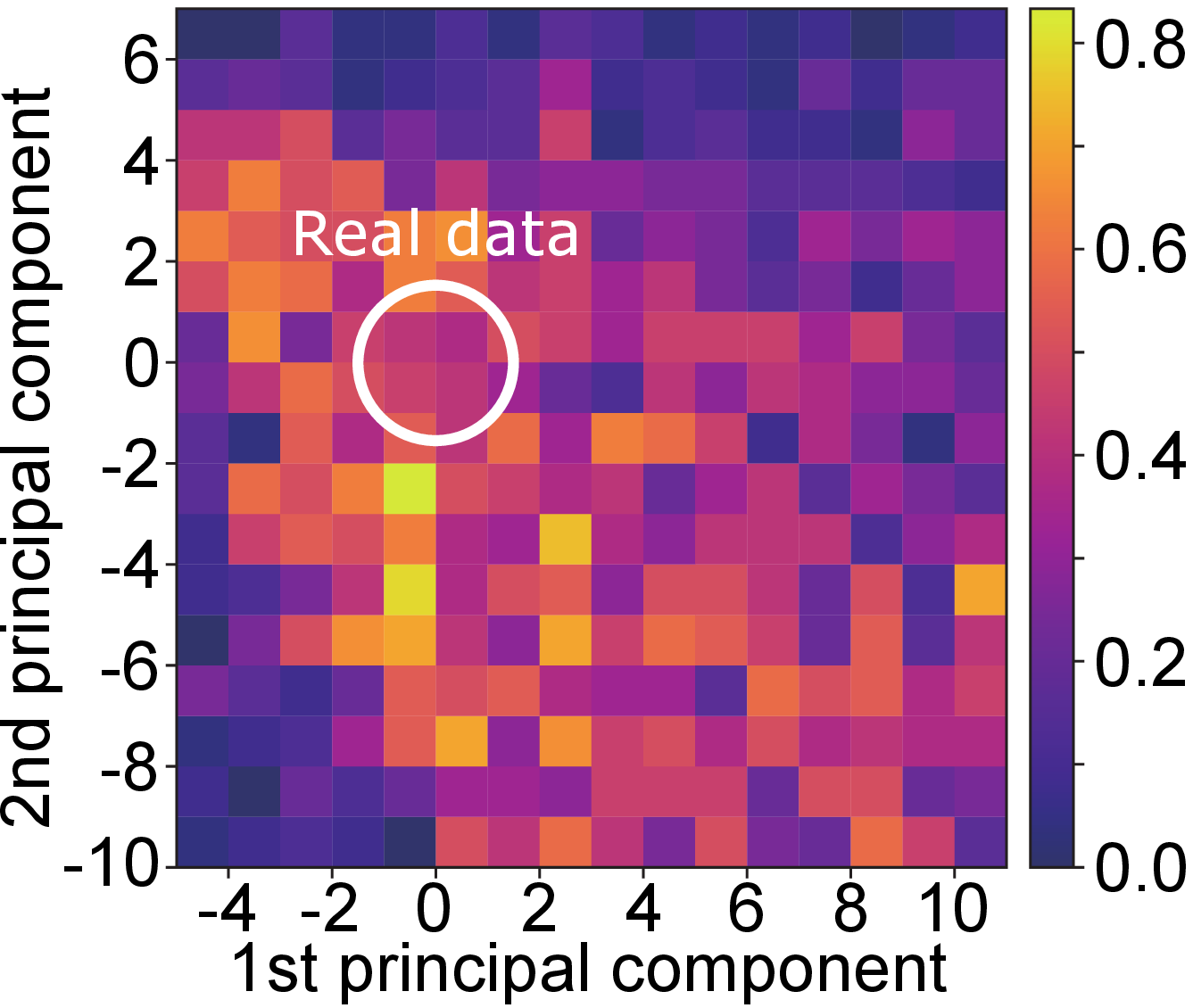}
        {Color map representing posterior probabilities in speech naturalness.}
        
        \drawfigwide{t}{0.75\linewidth}{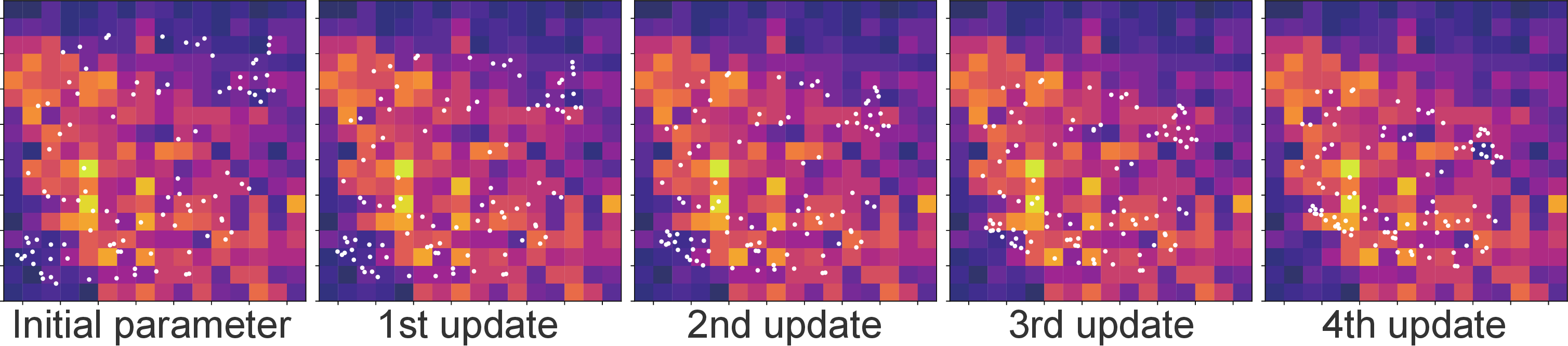}
        {Generated data (white point) at each training iteration. Color map shown in \Fig{posteriorprob} was also drawn but never used in training.}

        \drawfig{t}{0.75\linewidth}{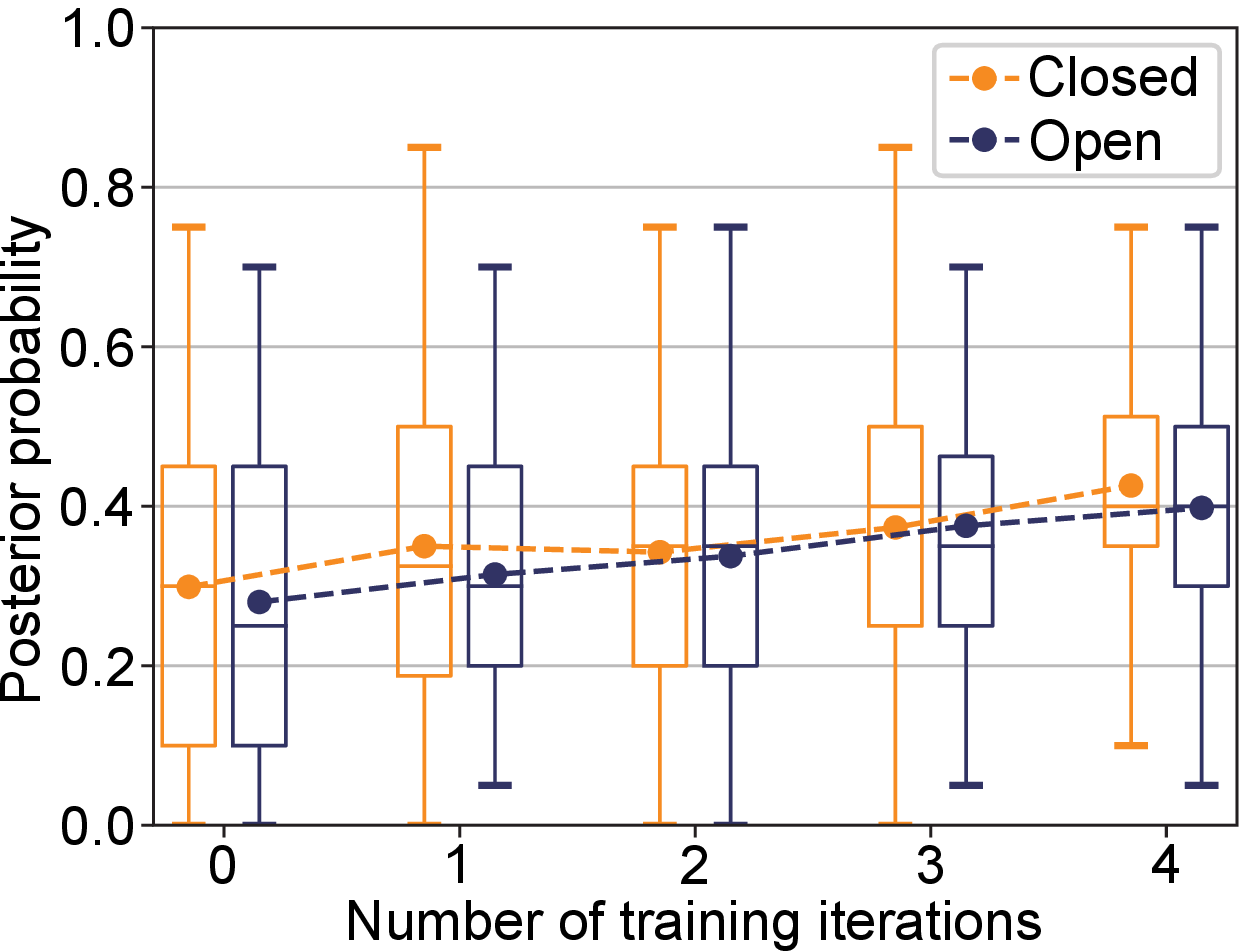}{Posterior probability at each iteration step. Box indicates first, second (i.e., median), and third quartiles. Line plot indicates mean value.}
    
        An experimental evaluation was done to investigate whether our HumanGAN can represent the perception distribution of speech naturalness, i.e., the HumanGAN tries to represent a distribution of speech features for which humans recognize the synthesized speech as human speech. Although the HumanGAN originally does not use real data, this evaluation used speech data for comparison with the basic GAN, dimensional reduction of speech features, and initialization of the generator. The speech data used was reading-style speech uttered by 199 female speakers included in the Vowel Database: Five Japanese Vowels of Males, Females, and Children Along with Relevant Physical Data (JVPD)~\cite{jvpd_corpus}. Before extracting speech features, the speech data were downsampled at $16$~kHz, and their powers were normalized. The speech features were $513$-dimensional log spectral envelopes extracted every $5$~ms. The WORLD vocoder~\cite{morise16world,morise16d4c} was used for the feature extraction. Julius~\cite{lee01julius} was adopted to obtain phoneme alignment, and we extracted speech features of the Japanese vowel /a/. We applied principal component analysis (PCA) to the log spectral envelopes of /a/ of 199 female speakers and extracted two-dimensional principal components. In this paper, we assumed that the real data of many speakers follows a two-dimensional standard Gaussian distribution, and we normalized the two-dimensional principal components to have zero-mean and unit-variance. For humans' perceptual evaluation, a speech waveform was synthesized from the generated speech features. First, the first and second principal components were generated from a generator and de-normalized. The remaining speech features, i.e., the third and upper principal components, the fundamental frequency, and aperiodicity components, were copied from one frame of an average-voice speaker. Average-voice speaker means a speaker whose her first and second principal components are the closest to 0. These correspond to speech features of one frame ($5$~ms). To make the perceptual evaluation easy for crowdworkers,we copied the features for $200$ frames and synthesized 1~s of a speech waveform by using the WORLD vocoder.          
    
    \subsection{Difference between real-data and perception distributions} \label{sec:diff}
        First, to emphasize the need for the HumanGAN, we demonstrate that a perception distribution is different from a real-data distribution in speech naturalness. We split the two-dimensional space into grids and evaluated humans' naturalness tolerance (i.e., posterior probability of speech naturalness). For the humans' evaluation, we instructed listeners to ``listen to a voice and answer with $1$ when perceiving the speech as very unnatural and $5$ when perceiving it as very natural'' and had listeners give a score on a 5-point scale (from 1 to 5). $1$-through-$5$ of the obtained scores corresponded to $0.00, 0.25, 0.50, 0.75, 1.00$ of the posterior probability $D\pt{\hatx_n}$. The evaluation was performed by using the Lancers crowdsourcing platform~\cite{lancers}. The posterior probability of each grid was set to the value averaged among listeners. At least five listeners scored for one grid, and the total number of listeners was $96$.
        
        \Fig{posteriorprob} shows the result. As described in \Sec{expcond}, the real data was normalized to have zero-mean and unit-variance, and the basic GAN represents this range. However, as shown in \Fig{posteriorprob}, the perception distribution covered a range wider than the real-data distribution. The basic GAN cannot represent this wider range, so the HumanGAN, which adopts perceptual evaluation by humans is needed for modeling the perception distribution. 
    
    \subsection{Transition of generated data during training} \label{sec:learnmodel}
        Next, to intuitively understand generator training, we ploted the data generated at each iteration step. The prior noise fed to the generator followed a two-dimensional uniform distribution $U(-1, 1)$. This prior noise was randomly generated before a training iteration and fixed during the iteration. The generator was a small feed-forward neural network consisting of a two-unit input layer, $2\times4$-unit sigmoid hidden layers, and two-unit linear output layer. The model parameters were randomly initialized, but we iterated the random initialization until the initial generator output data that covered ranges of a higher posterior probability, shown in \Fig{posteriorprob}. The gradient descent method with a learning rate of $\alpha = 0.0015$ was used for training. Chainer~\cite{chainerproc} was used for implementation. The number of generated data $N$, the number of perturbations, and the number of training iterations were set to $100$, $5$, and $4$, respectively. The standard deviation of NES was set to $1.0$. During the HumanGAN training, we instructed listeners to ``listen to two voices and answer with $1$ when perceiving the first one as significantly natural and $5$ when perceiving the second as significantly natural. Answer $3$ when perceiving the two samples as equally natural.'' The $1$-through-$5$ of the obtained scores corresponded to $1.0, 0.5, 0.0, -0.5, -1.0$ of the difference of the posterior probabilities, $\Delta D(\hatx_n^{(r)})$.
        
        \Fig{iteration} shows the data generated for each iteration step. The posterior probability of \Fig{posteriorprob} is also drawn, but note that we never used it in training. The figure indicates that the generated data transited from a lower probability range (darker area) to a higher probability range (brighter area) during the training iteration. Therefore, we can qualitatively demonstrate that the HumanGAN makes it possible to train a generator that represents a human's perception distribution. 
        
    \subsection{Increase of posterior probability in training}
        Finally, we confirmed that the training iteration increases the posterior probability. Here, we sampled new prior noise not used in training and generated new speech features. We call speech features generated from prior noise used in training ``closed data'' and those generated from prior noise not used in training ``open data.'' The posterior probabilities of the open data were scored by humans in the same manner as in \Sec{diff}. 

        \Fig{boxplot} shows a box plot at each iteration step. We can see that the training iteration increases the posterior probabilities of not only the closed data but also the open data. Therefore, we can quantitatively demonstrate that the generator of the HumanGAN can represent the perception distribution.

\vspace{-1mm} 
\section{Conclusion} \vspace{-1mm}
    This paper presented the HumanGAN, which can represent humans' perceptually recognizable distribution. The DNN-based discriminator of the basic GAN was replaced with human-based perceptual evaluation. The DNN-based generator of the HumanGAN was trained by using a black-boxed optimization algorithm and human evaluation. We evaluated the HumanGAN in speech naturalness modeling, and we qualitatively and quantitatively demonstrated that the HumanGAN can represent a humans' perceptually recognizable distribution. As future work, we address the scalability of the HumanGAN in terms of the number of data and feature dimensions.

\textbf{Acknowledgements:}
This research and development was supported by the SECOM Science and Technology Foundation and the MIC/SCOPE \#18210310.


\bibliographystyle{IEEEbib}
\bibliography{tts}

\end{document}